\documentclass{aa}
\usepackage{graphicx}
\usepackage{times}
\usepackage{natbib}
\usepackage{booktabs} 
\usepackage{morefloats}
\usepackage[hyphenbreaks]{breakurl}
\usepackage{amsmath}
\usepackage{amssymb}
\usepackage[noabbrev]{cleveref}

\def\apgt{\ {\raise-.5ex\hbox{$\buildrel>\over\sim$}}\ }
\def\aplt{\ {\raise-.5ex\hbox{$\buildrel<\over\sim$}}\ }
\def\lteq{\ {\raise-.5ex\hbox{$\buildrel<\over-$}}\ }

\def\aap{\ {A\&A}\ }

\def\apjl{\ {ApJL}\ }
\def\apjs{\ {ApJS}\ }

\def\mnras{\ {MNRAS}\ }

\def\pasj{\ {Publ. Astr. Soc. Japan}\ }

\begin{document}

\title{The formation of periodic three-body orbits for Newtonian systems}
\titlerunning{Forming braids}

\author{
Simon Portegies Zwart\inst{1}\thanks{None of the authors could have done this work without the other two.}
\and
Arjen Doelman\inst{2}
\and
Jelmer Sein\inst{3},
}

\institute{
Leiden Observatory, Leiden University, PO Box 9513, 2300 RA, Leiden, The Netherlands
\and
Mathematisch instituut, Leiden University, PO Box 9513, 2300 RA, Leiden, The Netherlands
\and
Niels Bohr International Academy, University of Copenhagen, Copenhagen, Denmark
}

\abstract
{Braids are periodic solutions to the general N-body problem in
gravitational dynamics. These solutions seem special and unique but may result from rather usual encounters between four bodies.
}
{We attempted to learn more about the existence of braids in the Galaxy by reverse engineering
the interactions in which they formed.}
{We carried out simulations of self-gravitating systems of N particles
using fourth-order integration. We started by constructing
the specific braid and subsequently bombard it with a single
object. We subsequently studied how frequently the bombarded braid
dissolves into four singles, a triple and a single, a binary and two
singles, or two binaries. The relative proportion of these events
gives us insight into how easy it is to generate a braid through the
reverse process.
}
{Braids are relatively easily generated from
  encounters between two binaries or between a triple and a single object, so
  long as they meet each other, independent of the braid's
  stability. We find that three of the explored braids are long-lived
  and linearly stable against small perturbations, whereas the fourth is
  unstable and short-lived. The shortest-lived braid appears to be the least
  stable and the most chaotic.  Nonplanar encounters can also lead to
  braid formation, which, in our experiments, are themselves planar.  The
  parameter space in azimuth and polar angle that leads to braid
  formation via binary--binary or triple--single encounters is
  anisotropic, and the distribution has a low fractal dimension.  }
{Since $\sim 9$\,\% of our calculations lead
  to periodic three-body systems, a substantial fraction, braids may be more common than
  currently assumed.  They could easily temporarily exist as a result of
  multi-body (binary--binary or triple--single) interactions. We do not
  expect many stable braids to exist for an extensive period of time,
  but they may be quite common as transients, surviving for tens
  to hundreds of periodic orbits. We argue that braids are
  particularly common in relatively shallow-potential background
  fields, such as the Oort cloud or the Galactic halo. The
  formation of braids, however, easily leads to collisions between two or more
  of their constituents. If composed of compact objects, they
  potentially form interesting targets for gravitational wave
  detectors. }


\maketitle

\section{Introduction}

The gravitational N-body problem originated when Newton presented his
theory of gravity \citep{Newton:1687}. Newton was able to analytically calculate periodic solutions of the two-body problem, which we call
"Kepler orbits" \citep{Kepler:1609}: two objects in a periodic bound
orbit in an elliptical trajectory around their common center of mass.
For $N>2$, periodic solutions exist, but they are hard to find.

Negative total energy alone cannot be used to
determine the long-term stability of a $\geq 3$ body problem because
the system is chaotic and therefore evolves in finite time to the
ejection of one object, leaving the other two bound in a pair
\citep{1891BuAsI...8...12P}. Euler and Lagrange, however, found
potentially long-lived periodic solutions to the three-body problem
\citep[see also][]{1976CeMec..13..267H}. These choreographies were
later identified as belonging to the same family of periodic solutions
\citep{0951-7715-11-2-011}. Although so far periodic solutions
have been shown to be possibly long-term-stable, there is no theorem to prove this
\citep{Terracini2022}.

More than a century later, \cite{1993PhRvL..70.3675M} proposed a
topological classification as a tool for extending the ``symbolic
dynamics'' approach to many-body dynamics. He described the Figure-8
solution (not to be confused with our Fig.~8, which shows the initial
azimuth and polar angle distribution for successful braid formation)
of the gravitational N-body problem as a braid. Braids are periodic,
reducible, and pseudo-Anosov orbits of the gravitational N-body
problem \citep{KAJIHARA2023108640}.  Such a braid emerges when you add
the same velocity vector to each of the bodies.  In the reference
frame, all bodies now travel as parallel strands through time while
each strand traces an orbit projected onto one spatial coordinate in
time, creating a braid.  The form of the Figure-8 then becomes
$(b_1b_2^{-1})^{3}$, where $b_i$ corresponds to the crossing of the
$i$-th strand over the $(i+1)$st \citep{1993PhRvL..70.3675M}. The
stability of such primitive braids was further studied by
\cite{Simo1994}, who argued that some have relative equilibrium
solutions. In astronomical terms, we would describe a braid as a
nonhierarchical periodic solution of the N-body problem.

\cite{simo2001new} made the distinction between linearly ``stable''
and ``unstable'' choreographies. Of the 18 tested, he found that only
the Figure-8 is stable. This finding is consistent with that of
\cite{1956AN....283...17S}.

\cite{Chenciner2002} made a distinction between the simple
choreography of the N-body problem, the definition of which is
``a periodic solution in which all N masses trace the same curve
without colliding,'' and the choreography of arbitrary complexity or
symmetry. Both types of choreography exist \citep{Chenciner2002}, although
their stability and frequency remain uncertain \citep{Simo2002}.
\cite{KapelaSimo2007} continued this work by studying the stability
of complex (asymmetric) braids composed of seven bodies. They
found that such a choreography can be dynamically stable.

The astro-dynamical community became aware of the work on braids
through the discussion by \cite{2000math.....11268C}, who further
addressed the Figure-8 periodic solution. They, however, did not
provide proof for the Figure-8's long-term stability.  Such proof was
provided in \cite{KapelaSimo2007}, who demonstrated the linear stability
of the Figure-8 orbit.

In the years since, many new braids of various types have been
found. Nowadays, most braids are discovered using
supercomputers. \cite{2018PASJ...70...64L} found 1223 new solutions,
which were later expanded with another 13315 by
\cite{2021SCPMA..6419511L}. We conjecture that there exist an infinite
number of possible (distinct) periodic solutions to the general N-body
problem. Only a subset are expected to be stable against small
perturbations, but even so we still expect an infinite number of
stable periodic orbits.

Although \cite{2021SCPMA..6419511L} assert that these braids are stable, it is not
a priori clear to what degree they are long-term-stable.
The lack of stable braids does not necessarily mean that they do not
exist in the Universe but rather that it may be nontrivial to
recognize them. We followed up on the work of
\cite{2000MNRAS.318L..61H}, who considered the possibility that
Figure-8 braids form from two encountering pairs of objects. In this
case, objects could be black holes, stars, planets, planetesimals, or
any set of massive objects.

For our calculations, we adopted Newton's equations of motion with
point masses and integrated using a fourth-order numerical scheme. This
poses some limitations to the interpretation of our results, as
astronomical objects tend to have properties that cause them to
interact non-Newtonianly, such as finite radii, orbits
affected by general relativistic effects, and radiation
pressure\footnote{We followed Orwell's "Ignorance is strength"
phlosophy in this perspective \cite[][see p.6]{Orwell1949}.}.  The
effect of numerical errors, and the accuracy in relation to the chaotic
dynamics, is discussed in Sect. \ref{sect:accuracy}.

\section{Initial conditions and model parameters}

\subsection{Initial conditions}

We adopted a similar approach as proposed in
\cite{2000MNRAS.318L..61H}, by integrating a three-body braid and
bombarding it with a fourth particle. Since Newton's equations of
motion are time symmetric, the inverse interaction then results in a
braid, while the original bomb escapes. \cite{2000MNRAS.318L..61H}
used this strategy to explain how the Figure-8 braid can form from two
interacting binaries. We took a more general approach, covering a
larger part of the initial parameter space. In the main part of our
study, we reduced the initial parameter space by limiting ourselves to
interactions in the $x-y$ plane, setting the $z$ coordinate and the
velocity perpendicular to the plane equal to zero (note that it is not
known if nonplanar braids exist).  We further fixed the braid's
orbital phase and orientation but allowed the incoming object to come
from any azimuthal direction $\theta$. We randomized in this
direction. We relax the assumption on planar encounters in
Sect. \ref{sect:isotropic_encounters}, where we discuss the anisotropy
of the initial parameter space.

\subsection{Parametrics}\label{Sect:parameters}

Once the braid (see Fig. \ref{fig:braids}) is specified, the mass $m_4$,
velocity $v_4$, impact parameter $d$, and initial distance $r$ of the
incoming fourth object remain free to vary.  For initial conditions we
then have $v_4(t=0) \equiv v_{\rm in}$, $m_4(t=0) \equiv m_{\rm in}$,
and $d_4(t=0) \equiv d_{\rm in}$.  Note that if the velocity at
infinity satisfies $v_{\rm in} = 0$, the impact parameter $d_{\rm in}$
becomes meaningless. Here units are dimensionless, adopting N-body
units following Henon-Heggie, i.e., $G = 1$
\citep{1986LNP...267..233H}.

Simulations were continued until one object (or the center of mass of
a binary or triple) satisfied the following conditions: it is farther
than $1000$ N-body units from the center of mass of the four-body
system and is receding from the system with a velocity exceeding the
escape speed.  It is worth noting that the braid can formally be
unbound if an incoming object's kinetic energy exceeds the braid's
binding energy.  For the braids in our experiments, this is the case
if the incoming object has mass $m_{\rm in} = 1$ with velocity $v_{\rm
  in} \gtrsim 1.7$. An incoming object with even higher velocity can
unbind the braid into four single objects. Unbinding the braid,
however, requires efficient energy transfer from the incoming object
to the particles in the braid, which is not common (see Table
\ref{tab:Lyapunov_timescale}).  Table \ref{tab:Lyapunov_timescale}
also lists the total number of calculations performed, and the
fractions of the various resulting configurations.

\subsection{Classifying braids}

Apart from the Figure-8 braid, we studied two more complex braids,
referred to in \cite{2018PASJ...70...64L} as $I.A.^{i.c.}_4(0.5)$, and
$I.A.^{i.c.}_{68}(0.5)$. These names stand for the class of braid,
with in parentheses the mass of the third particle ($m_3$) assuming
$m_1=m_2=1$. The initial conditions for these braids can be written as
follows: ${\vec r}_1 = (-1, 0) = -{\vec r}_2$. ${\vec v}_1 = (\xi_a,
\xi_b) = -{\vec v}_2$. Here ${\vec r}_i \equiv (x_i, y_i)$ is the
two-dimensional Cartesian position vector, and equivalently for ${\vec
  v}_i$. The two scalars $\xi_a$, and $\xi_b$, characterize the
interaction. For the incoming third particle ${\vec r}_3 = (0, 0)$,
and ${\vec v}_3 = (2\xi_1/m_3, -2\xi_2/m3)$. All $z$ coordinates and
velocities $v_z$ are zero (but see Sect. \ref{sect:isotropic_encounters}).

We included a fourth braid with unequal masses, $m_1=0.87$, $m_2=0.80$,
and $m_3=1$, which is presented in Table\,1 of
\cite{2021SCPMA..6419511L}. The initial conditions for the braids we
explored are presented in Table \ref{tab:Initial_conditions}.

\begin{table*}
  \begin{center}
  \caption{The four braids used in this study.}.
\begin{tabular}{|l|ll|ll|}
\hline
$m$ & $x$ & $y$ & $v_x$ & $v_y$ \\
\multicolumn{5}{|l|}{Figure-8 problem (top panel in Fig. \ref{fig:braids})} \\
\hline
1 & 0.9700436 & -0.24308753 & 0.466203685 & 0.43236573 \\
1 & -0.9700436 & 0.24308753 & 0.466203685 & 0.43236573 \\
1 & 0 & 0 & 0.93240737 & 0.86473146 \\
\hline
\multicolumn{5}{|l|}{Non-equal-mass problem ($=$ model A, second panel in Fig. \ref{fig:braids})} \\
\hline
0.87 & -0.1855174644 & 0 & 0 & 2.0221546880 \\
0.80 & 1 & 0 & 0 & 0.3968976468 \\
1.0 & 0 & 0 & -2.25026390304 & 0 \\
\hline
\multicolumn{5}{|l|}{AI4(0.5) (third panel in Fig. \ref{fig:braids})} \\
\hline
1 & -1 & 0 & 0.2009656237 & 0.2431076328 \\
1 & 1 & 0 & 0.2009656237 & 0.2431076328 \\
0.5 & 0 & 0 & -0.8038624948 & -0.9724305312 \\
\hline
\multicolumn{5}{|l|}{AI68(0.5) (bottom panel in Fig. \ref{fig:braids})} \\
\hline
1 & -1 & 0 & 0.2138410831 & 0.0542938396 \\
1 & 1 & 0 & 0.2138410831 & 0.0542938396 \\
0.5 & 0 & 0 & -0.8553643324 & -0.2171753584 \\
\hline
\end{tabular}
 \tablefoot{The Figure-8, model A,
  $I.A.^{i.c.}_4(0.5)$, and $I.A.^{i.c.}_{68}(0.5)$.  Presented for
  these braids are the mass of the three components $m$, the Cartesian
  positions $x$, and $y$, and the corresponding velocities $v_x$, and
  $v_y$.  For all particles the Cartesian coordinate $z=0$, and
  velocity $v_z=0$.
\label{tab:Initial_conditions}
}
\end{center}
\end{table*}

\section{Numerical integration}

\begin{figure}
\center
\includegraphics[width=0.8\columnwidth]{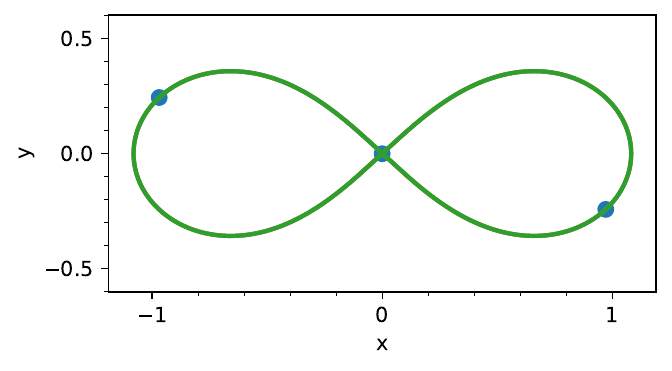}
~\includegraphics[width=0.8\columnwidth]{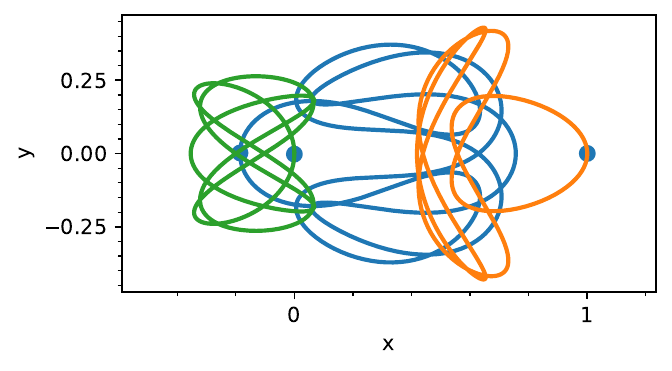}
\includegraphics[width=0.8\columnwidth]{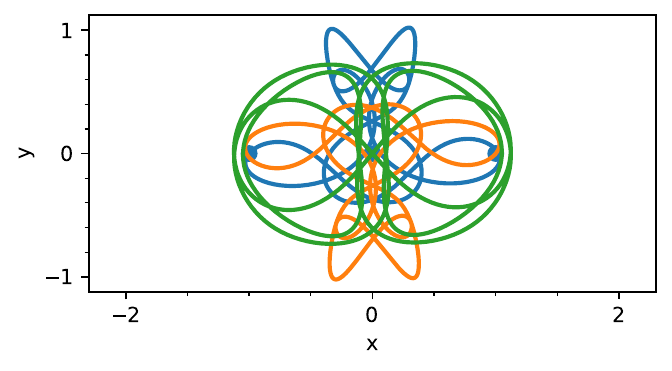}
~\includegraphics[width=0.8\columnwidth]{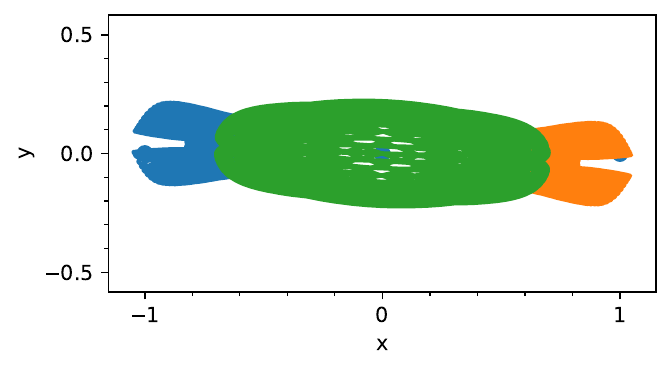}
\caption{The four three-body braids adopted in this study: the classic
  Figure-8 problem (\textit{top}) and three more complicated braids,
  the unequal-mass problem, $I.A.^{i.c.}_4(0.5)$, and
  $I.A.^{i.c.}_{68}(0.5)$ (\textit{bottom}). They are integrated with
  a time-step parameter of $\eta = 0.01$ with output every $0.001$
  time units for one periodic orbit. The $I.A.^{i.c.}_{68}(0.5)$ braid
  is a complex orbit for which the period exceeds 83 time units, much
  more than any of the others. The orbit closes only after many
  orbits, giving rise to the filled-in surfaces. For the Figure-8,
  only one line color is visible because all three objects have the
  same orbit and therefore overlap.
\label{fig:braids}
}
\end{figure}

We performed the numerical integration using the Astrophysics
Multipurpose Software Environment (AMUSE; \citealt{2009NewA...14..369P,2018araa.book.....P}) with the direct fourth-order Hermite predictor-corrector N-body solver {\tt Ph4}
\citep{2012ASPC..453..129M,2022A&A...659A..86P}. The calculations are
performed using point masses with a time-step parameter of $\eta =
0.01$, and zero softening. In Fig. \ref{fig:braids} we illustrate the
braids integrated for one orbit, and in Sect. \ref{sect:accuracy} we
discuss the effect of numerical precision on the results.

\begin{figure}
\center
\includegraphics[width=0.8\columnwidth]{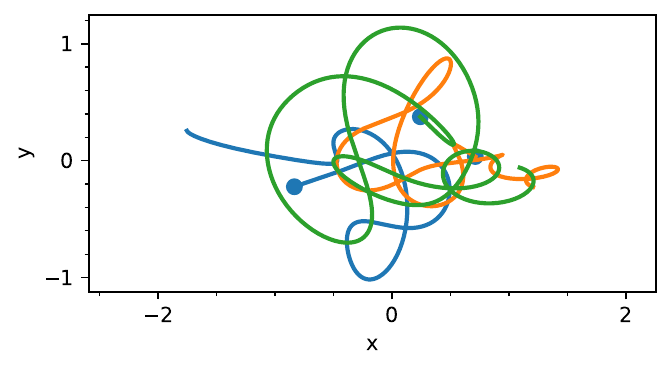}
\includegraphics[width=0.8\columnwidth]{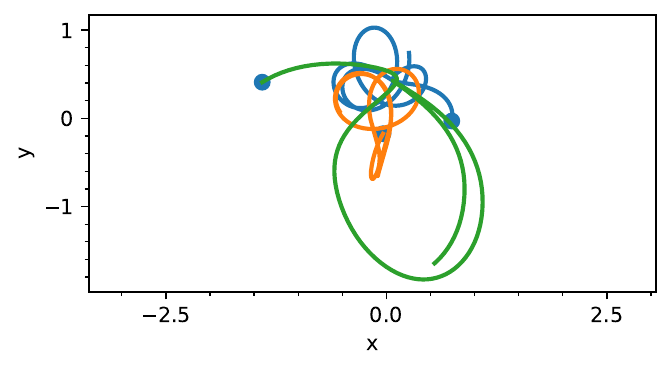}
\caption{Two representations of braid $I.A.^{i.c.}_4(0.5)$ after 100
time units. \textit{Top}: Unperturbed solution integrated
for ten N-body time units. \textit{Bottom}: Result of the
perturbed solution in the same time frame. The introduced
perturbation was $10^{-5}$ in the $x$ coordinate. Both solutions
have clearly deviated from the original braid, and from
each other. By this time, the phase space distance between the two
solutions is on the order of unity. The orbits are unstable, and in the
next 100 crossing times, one particle is ejected.
\label{fig:braidiA4m05_after_time_100}
}
\end{figure}

In Fig. \ref{fig:braidiA4m05_after_time_100} we present
$I.A.^{i.c.}_4(0.5)$ after 100 N-body time units. By this time, the
orbit cannot be classified as a braid; the orbit is not periodic. In
the next few hundred time units, one of the three particles is ejected,
leaving the other two orbiting their joined center of mass.

\begin{figure}
\center
\includegraphics[width=0.8\columnwidth]{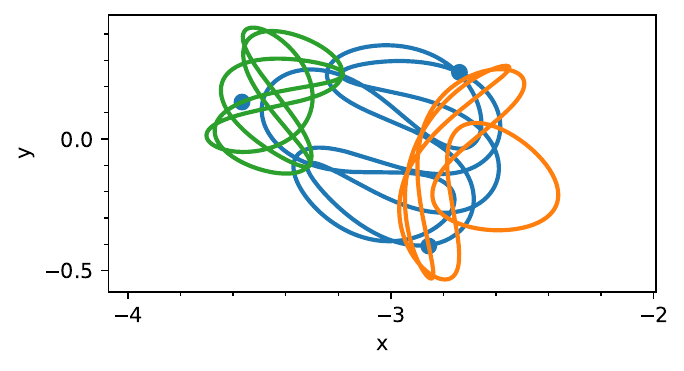}
\caption{Braid model A after $10^6$ N-body time units. By this time,
an initial perturbation of $10^{-5}$ has grown to {\cal
O}($10^{-2}$). The braid still resembles the original,
but the entire orbit has been precessing.
\label{fig:ModelA_after_time_10e6}
}
\end{figure}

Figure \ref{fig:ModelA_after_time_10e6} shows a distinctly different
evolution for model A (the unequal-mass braid case) after $10^6$ time
units.  By this time, the orbit is still stable, and the orbital shape
is indistinguishable from the original orbit, except that it has been
subject to precession. We integrated the orbit twice, once with a
perturbation of $10^{-5}$ to one of the particle's
x-coordinates. However, upon optical inspection, the two orbits are still
indistinguishable after integrating for $t = 10^6$.

\section{Results}

\subsection{Braid stability}\label{Sect:Braid_stability}

We determined the stability of a braid by integrating it twice, once
with the canonical initial realization (see
Table \ref{tab:Initial_conditions}) and once with a small perturbation by
translating one particle $10^{-5}$ along the x-axis. Both integrations
were continued for $10^5$ N-body time units, and we determined the
Lyapunov timescale by fitting a straight line to the logarithm of the
geometric distance evolution in time. The results are presented in
Table \ref{tab:Lyapunov_timescale}.

The phase-space distance of an initially introduced perturbation in
$I.A.^{i.c.}_4(0.5)$ grows exponentially, as one would expect for a
chaotic system. The Lyapunov timescale for this braid is $\sim 13$
orbits. The other models, Figure-8, model A, and braid
$I.A.^{i.c.}_{68}(0.5)$, are linearly stable against infinitesimal
perturbations. Note that for braid $I.A.^{i.c.}_{68}(0.5)$ $t_{\rm stable}
\apgt 9.10^4$ (see Table \ref{tab:Lyapunov_timescale}), which strictly
speaking does not make it linearly stable, but the accumulation errors
in our numerical integration may drive the system's
instability here. Such numerical errors can be mitigated by integrating the
braid using arbitrary precision arithmetic, as is explored in
\cite{2015ComAC...2....2B}, which is beyond this papers' scope.

Interestingly enough, the growth for model A makes periodic
excursions, to shrink to values comparable to the initially introduced
perturbation ($10^{-5}$). The timescale of these excursions is about
$\sim 3850$ time units, or 642 orbits. This curious behavior can be
interpreted as the system being stable against small variations in
orbital phase. Even after integrating the system for a million time
units it still exhibits the same curious behavior. This corresponds
roughly to a 5-minute variation in the Earth's orbit around the Sun.
Such quasi-periodic behavior of excursions in shrinking and
stabilizing, with a well-defined timescale, is a hallmark signature
of a Neimark-Sacker bifurcation -- the discrete time analog of the
Hopf bifurcation \citep{homburg2024bifurcation}.

\subsection{Braid formation}

\begin{table*}
\caption{Integration result for the various braids.}
\begin{tabular}{|l|rrrrr|llll|l|}
\hline
Model & $E$ & $t_{\rm peri}$ & $t_L$ & $t_{\rm stable}$ & $N$ & $f_{BSS}$ & $f_{4S}$ & $f_{BB}$ & $f_{TS}$ & stability\\
\hline
Figure-8 &-1.29&6.33 & $\aplt 3 \cdot 10^4$& $>10^5$ & 4380& 0.833 & 0.079 & 0.064 & 0.025 & linear stable \\
Model A &-2.08&5.99 & $\apgt 3 \cdot 10^4$ & $>10^5$ & 4175 & 0.817 & 0.078 & 0.083 & 0.022 & linear stable \\
$I.A.^{i.c.}_4(0.5)$ &-1.00 &19.01 & $\aplt 2 \cdot10^2$ & $\aplt 40$& 3624 & 0.877 & 0.055 & 0.037 & 0.031 & unstable \\
$I.A.^{i.c.}_{68}(0.5)$ &-1.26&83.85 &$\sim 10^4$& $\aplt 9\cdot 10^4$& 3020 & 0.803 & 0.094 & 0.064 & 0.039 & marginally stable\\
\hline
\end{tabular}
 \tablefoot{Integrations were
  performed using the fourth-order Hermite predictor corrector for $10^4$
  N-body time units with a time-step parameter of $\eta=0.01$, and
  with outputs every 0.1 time unit. The first column identifies the
  interaction, followed by the braid's total energy in N-body units
  [mass$\cdot$(length/time)$^{2}$], and the period of one complete
  orbit. The Lyapunov timescale (fourth column) was calculated by
  introducing a small perturbation (of $10^{-5}$) in the x-coordinate
  of the first particle. The fifth column gives $t_{\rm stable}$, the
  integration time over which we could determine that the system
  remained stable with a maximum integration time of $t_{\rm stable} =
  10^4$.  Note that we classify $I.A.^{i.c.}_{68}(0.5)$ as marginally
  stable, discussed in Sect. \ref{Sect:Braid_stability}.  The following
  column gives the number of experiments calculated while varying
  $m_{\rm in}$, the impact parameter and relative encounter
  velocity. The last 4 columns give the fraction of interaction that
  leads to the various outcomes (marginalized over $v_{\rm in}$,
  $m_{\rm in}$, and $d_{\rm in}$), binary-single-single (BSS), four
  single stars (4S), binary-binary (BB), and triple-single (TS).
\label{tab:Lyapunov_timescale}
}
\end{table*}

The majority of interactions result in a binary with two single
objects, while the other three possibilities have comparable
probabilities. (Note that interactions leading to four single stars, 4S
interactions, only occur for sufficiently large incoming velocities.)
The relative branching ratios of the encounter results are quite
comparable for each of our selected braids, and this measure appears
independent of the braid's stability. On average, 4 to 6\,\% of the
braids fall apart to a binary-binary, consistent with the earlier
results by \cite{2000MNRAS.318L..61H} for the Figure-8 problem. We
also find a somewhat lower fraction of braids (of $\sim 2$ to $\sim
4$\,\%) that dissolve into a triple-single. Both these solutions are
of interest for generating braids from the reverse interaction.

\begin{figure}
\center
\includegraphics[width=0.8\columnwidth]{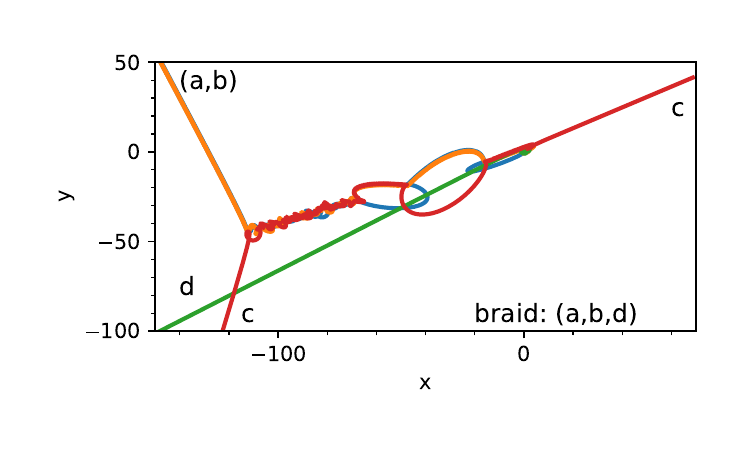}
\includegraphics[width=0.8\columnwidth]{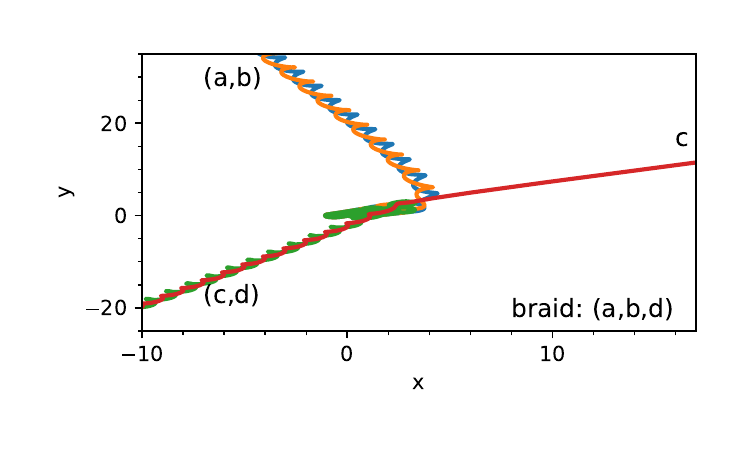}
\includegraphics[width=0.8\columnwidth]{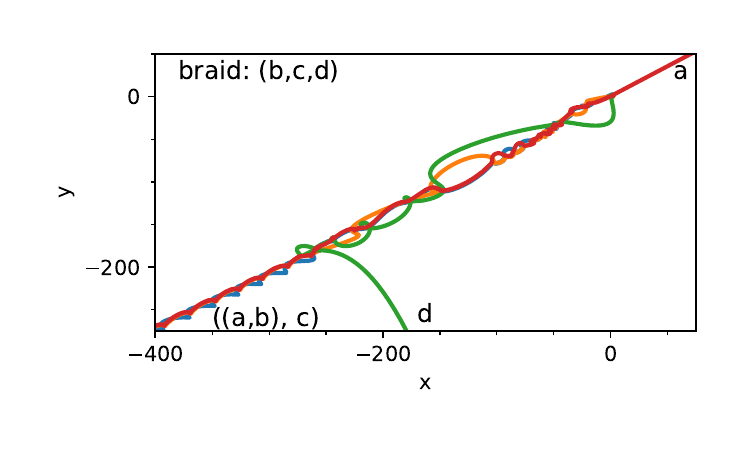}
\caption{Three examples of how the braid $I.A.^{i.c.}_4(0.5)$ forms
from an encounter between a binary (a,b) and two single stars (c
and d), between two binaries (a,b and c,d), and a triple
(a,b,c) encountering a single (d).
\label{fig:interaction}
}
\end{figure}

In Fig. \ref{fig:interaction} we present the three most common
interactions that lead to braid-formation, which are a braid
$I.A.^{i.c.}_4(0.5)$ by bombarding a binary with two single stars, two
binaries that encounter each other to make the Figure-8 (as in
\citealt{2000MNRAS.318L..61H}), or forming a braid $I.A.^{i.c.}_4(0.5)$
by a single object injected into a triple.

\subsection{Braid progenitors}

\begin{figure}
\center
\includegraphics[width=1\columnwidth]{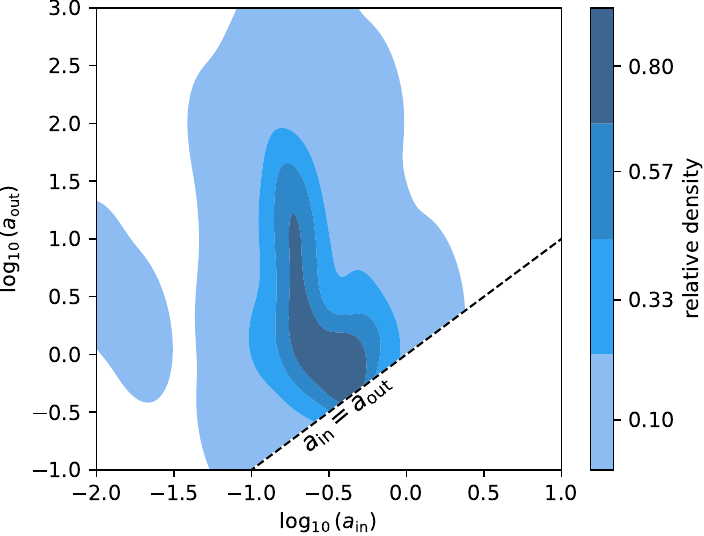}
\includegraphics[width=1\columnwidth]{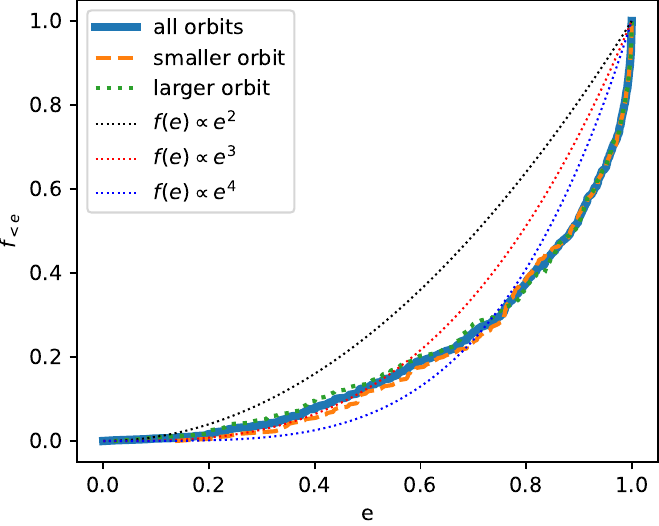}
\caption{Model A distributions of orbital elements, semimajor axes,
and eccentricities for an interaction resulting in a
binary--binary with $f_{\rm BB} = 0.058$ (see
Table \ref{tab:Lyapunov_timescale}). \textit{Top}: Semimajor
axis of the two binaries. We present the tighter binary along
the x-axis and the wider binary along the y-axis (explaining the
empty lower-right corner bordered by the dashed line). \textit{Bottom}: Eccentricity distributions of various orbits. The
Kolmogorov-Smirnov test indicates that for model A, the tighter and
wider distributions for eccentricity are indistinguishable. For the
Figure-8 ($p=0.0064$) and $I.A.^{i.c.}_{68}(0.5)$
($p=0.034$), the eccentricity distribution of the tighter binaries
appears not to be randomly sampled from a single parent
distribution.
\label{fig:orbital_elements}
}
\end{figure}

In Fig. \ref{fig:orbital_elements} we present the orbital separations of
the two binaries for braid model A, the one with a $f_{BB}=0.83$ in
Table \ref{tab:Lyapunov_timescale} or 242 binary pairs. The inner and outer
binaries have $\langle a_{\rm in} \rangle \simeq 0.30$ with a median
of $a_{\rm in} = 0.24^{+0.47}_{-0.14}$. For the outer orbit we find
$\langle a_{\rm out} = \rangle \simeq 52$ with a median of $a_{\rm
out} = 1.84^{+21.34}_{-0.46}$. For the other braids, the media have a
similar range. It turns out that braids tend to dissolve in a very
tight and one very wide binary.The braids also tend to dissolve
in a hierarchical triple (one with a tight inner a rather wide outer
orbit).

The eccentricity ($e$) distribution of the two binaries in which a
braid separates is strongly peaked toward unity. Objects in our
calculations were point masses, and collisions or gravitational wave
radiation were ignored. But if we were to introduce (reasonable)
finite radii, $\apgt 10$\% of interactions would lead to collisions,
rather than a stable pair of binaries. To guide the eye, and
illustrate the skewness of the eccentricity distribution, we overplot
curves for $f(e) \propto e^2$, $\propto e^3$, and $\propto e^4$, in
Fig. \ref{fig:orbital_elements}. A distribution $f(e) \propto e^2$ would
naively be consistent with expectations based on virialized few-body
scattering \citep{1975MNRAS.173..729H}. However, a steeper distribution
toward high eccentricities is expected for encounters between hard
pairs of particles \citep{2024MNRAS.531..739G}. It raises the
intriguing possibility that a braid could form from a four-body encounter
in which two objects experience a collision, leaving the collision
product and the other unaffected objects in a braid.

\subsection{Braid formation cross section}

\begin{figure}
\center
\includegraphics[width=1\columnwidth]{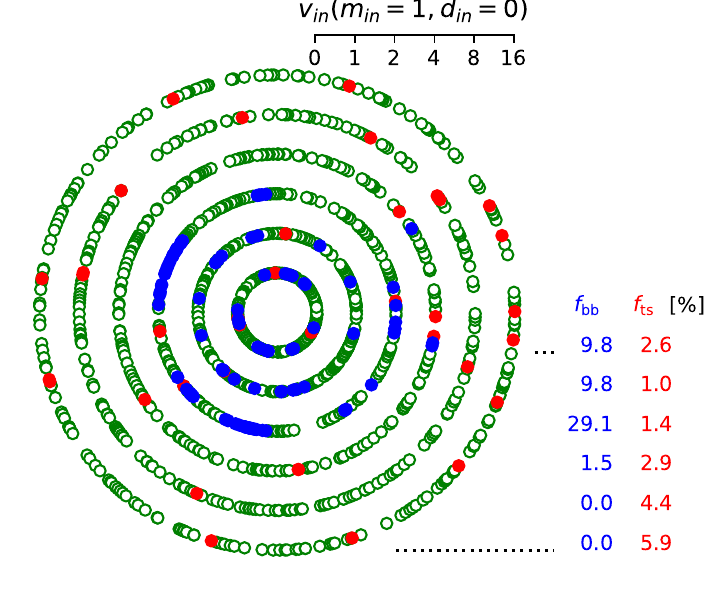}
\includegraphics[width=1\columnwidth]{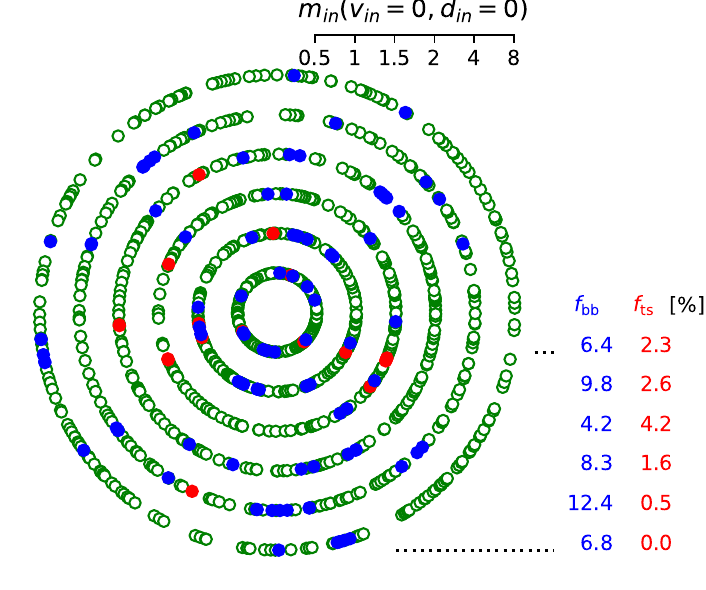}
\includegraphics[width=1\columnwidth]{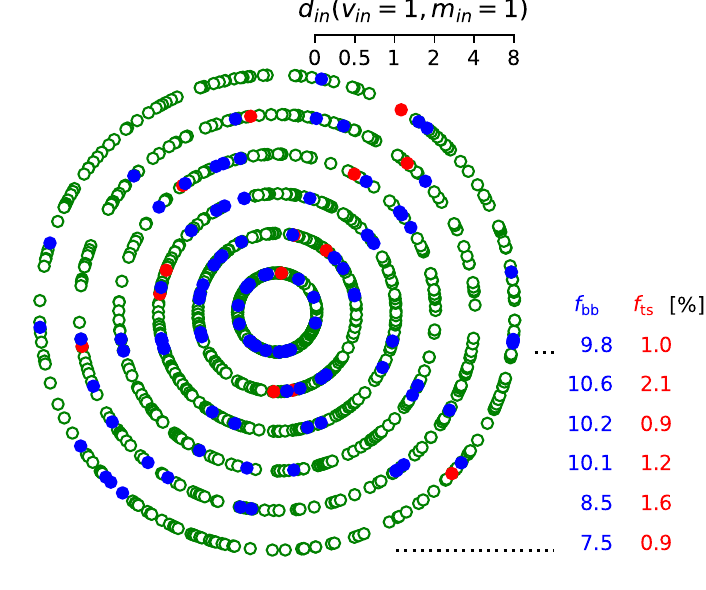}
\caption{Result of the encounter with model A when varying $v_{\rm
in}$ (\textit{top}), $m_4$ (\textit{middle}), and $d_{\rm in}$ (\textit{bottom})
for six values of the free parameter. The results are presented along
the circle $\phi$, and the parameter values corresponding to the
concentric circles are indicated along the scale bar on the top
middle to right of each figure. To the bottom right, we show the
fraction of interactions that resulted in a binary--binary (blue)
and a triple--single (red). These colors also correspond to the colors
of the bullets in the concentric circles, and open green
circles indicate the most common result of a binary with two singles.
For comparison, the circle in the top panel is identical to the second
(from the inside) circle in the middle panel. The second circle in
the top panel is identical to the inner circle in the bottom panel.
\label{fig:ModelA_results}
}
\end{figure}

\begin{figure}
\center
\includegraphics[width=1\columnwidth]{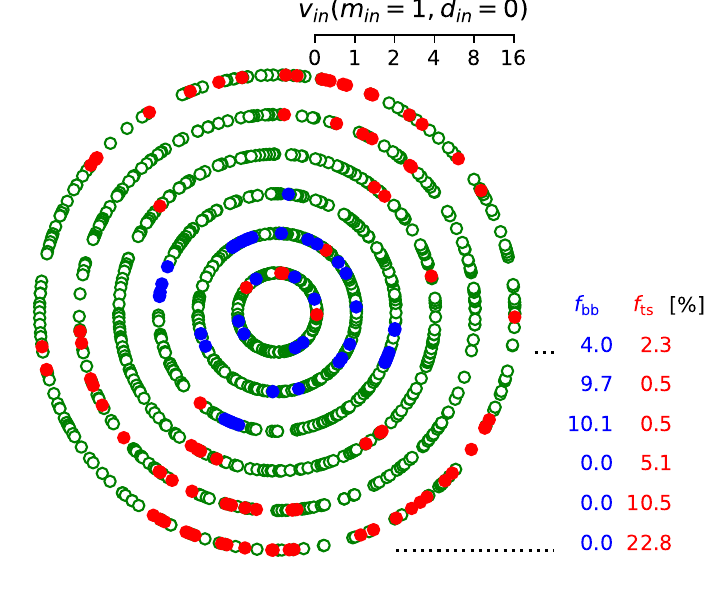}
\caption{Same as Fig. \ref{fig:ModelA_results} but for braid $I.A.^{i.c.}_{68}(0.5)$.
\label{fig:IA68_results}
}
\end{figure}

In Fig. \ref{fig:ModelA_results} we present the interaction result
from launching particle $m_4(t=0) \equiv m_{\rm in}$, with velocity
$v_{\rm in}$, and impact parameter $d_{\rm in}$ into the braid from
model A. We performed six series of calculations for each parameter,
marginalizing over the direction ($\phi$) from which the particle was
injected. We performed similar calculations for the other braids, but
the figures are insufficiently different to present. Except for the
effect of $v_{\rm in}$ for braid $I.A.^{i.c.}_{68}(0.5)$, which we
present in Fig. \ref{fig:IA68_results}, showing quite different
results. It is interesting to note that, together with the Figure-8,
this braid is linearly stable.

The circles with outcome statistics in Fig. \ref{fig:ModelA_results}
illustrates irregularity in the solution space. Although outcomes are
dominated by binary and two single escapers (green circles), the
binary-binary formation channel and triple-single channel are
prevalent. For this model, when the incoming velocity $v_4 \apgt 4$,
encounters fail to result in two binaries.

On the other hand, even higher incoming velocities tend to lead to the
survival of the braid, rather than to the formation of a hierarchical
triple.  In Fig. \ref{fig:ModelA_results} we do not make the distinction
between a surviving braid (democratic triple) and the more usual
outcome in the form of a hierarchical triple; both are indicated by
red dots in Fig. \ref{fig:ModelA_results}. This trend is even more
pronounced in Fig. \ref{fig:IA68_results}, where the braids in the
outermost ring (highest incoming velocity) survives in 22.8\% of the
cases. When injected along the braid's long axis, the encounter tends
to lead to a binary and two singles, or four singles. In
Fig. \ref{fig:IA68_results} we present the simulation results for braid
$I.A.^{i.c.}_{68}(0.5)$ with ${\vec v}_4$ as the free parameter. One
of the striking differences compared to Fig. \ref{fig:ModelA_results} is
the relatively large proportion of triples formed when the velocity of
the incoming particle, $|{\vec v}_4| \apgt 8$.

However, when injected perpendicular to the braid's long axis (around
$45^\circ \aplt \phi \aplt 135^\circ$ and $225^\circ \aplt \phi \aplt
315^\circ$) the encountering objects tend to fly through the braid
without doing much damage. The other models (Figure-8, model A, and
$I.A.^{i.c.}_4(0.5)$), being more symmetric, are less prone to such
harmless penetrating encounters. It is interesting to note that the
survivability of the braid (or the preservation of the democratic
triple) as a result of a penetrating encounter seems not to depend on
stability, or the Lyapunov timescale.

\subsection{The third dimension}\label{sect:isotropic_encounters}

\begin{figure}
\center
\includegraphics[width=1\columnwidth]{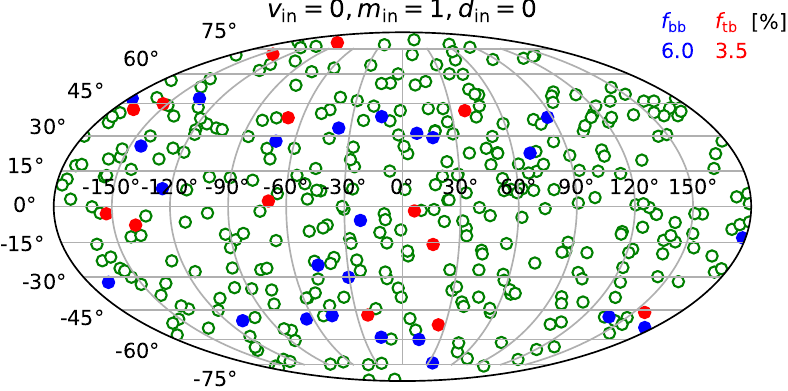}
\includegraphics[width=1\columnwidth]{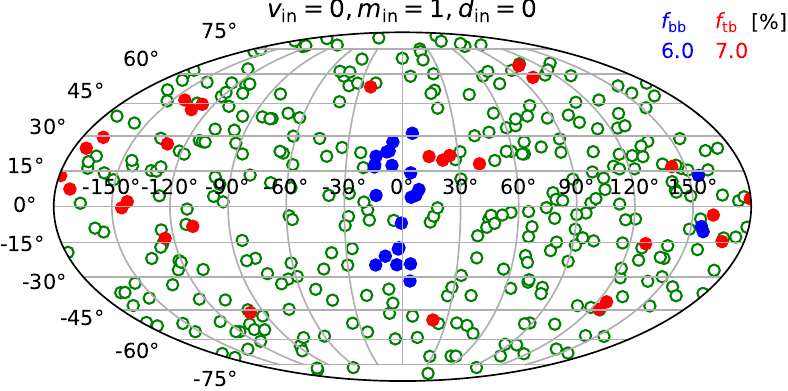}
\caption{Mollweide projection (azimuth along the x-axis and polar
  angle) for encounters of the Figure-8 braid with $v_{\rm in} = 0$
  (\textit{top}) and $v_{\rm in} = 2$ (\textit{bottom}).  The other
  parameters are $m=1$ and $d=0$.  Each simulation was run 300 times with
  $\eta = 0.01$.  For the legend of the symbols, see
  Fig. \ref{fig:ModelA_results}.
\label{fig:Mollweide_projection}
}
\end{figure}

So far we have examined the planar problem, in which all participating
particles are in the same plane.  Since this is rather restricted, we
also explored the possibility forming a planar braid given a nonplanar
interaction. We realized this by injecting the incoming particle from a
random direction into the planar braid. In this experiment the braid
is orbiting in the X-Y Cartesian plane with zero azimuth ($\theta =
0$). The incoming particle is injected isotropically from azimuth
$\theta \in [-180^\circ, 180^\circ]$ and polar angle $\phi \in
[-90^\circ, 90^\circ]$. In Fig. \ref{fig:Mollweide_projection} we present
the Mollweide projection of one such cases for the Figure-8 problem
for two incoming velocities: $v_{\rm in} = 0$, and $v_{\rm in} = 2$.

The fraction of interactions that produce braids in the nonplanar
case is statistically indistinguishable from that in the planar case,
for binary-binary as well as for the triple-single encounters.

Certain areas in parameter space are preferred for specific
interactions.  This trend becomes even more pronounced for higher
initial incoming velocities.  For the Figure-8 solutions (adopting
$m_{\rm in} = 1$, and $d_{\rm in} = 0$) braids, a low value of $v_{\rm
  in} = 0$ does not lead to an appreciable anisotropy in the initial
conditions for forming braids either from binary-binary or
triple-single encounters. However, as illustrated in
Fig. \ref{fig:Mollweide_projection}, for $v_{\rm in} = 2$ binary-binary
encounters are significantly more concentrated to low values of the
azimuth and polar angles.  This trend of clustering is comparably
pronounced in the model A, and $I.A.^{i.c.}_{68}(0.5)$ braids, but is
notoriously absent for $I.A.^{i.c.}_4(0.5)$, which does not show this

\begin{table}
\caption{Chi-squared values for the Rayleigh test.}
  \begin{center}
\begin{tabular}{|l|lll|lll|}
\hline
                     & \multicolumn{3}{|c}{$v_{\rm in}=0$} & \multicolumn{3}{|c|}{$v_{\rm in}=2$} \\
Model                & $R_{\rm BSS}$ & $R_{\rm BB}$& $R_{\rm TS}$ & $R_{\rm BSS}$ & $R_{\rm BB}$& $R_{\rm TS}$ \\
\hline
Figure-8             & 0.4 & 2.7 & 0.6  &  4.3 & 15.7 & 4.9 \\
Model A              & 1.5 & 1.6 & 0.8  &  5.8 & 12.1 & 1.2 \\
$I.A.^{i.c.}_4(0.5)$   & 1.6 & 0.3 & 3.4  &  0.1 & 0.6 & 0.8 \\
$I.A.^{i.c.}_{68}(0.5)$ & 2.1 & 1.8 & 3.1  & 1.1  & 6.1 & 2.5 \\
\hline
\end{tabular}
  \tablefoot{Statistical tests with
  three degrees of freedom for spherical isotropy in azimuth and polar
  angles for the various simulation outcomes for braids Figure-8,
  model A, $I.A.^{i.c.}_4(0.5)$, and $I.A.^{i.c.}_{68}(0.5)$. We
  performed 300 simulations for each set of initial conditions. In both
  sets $m_{\rm in}=1$ and $d_{\rm in}=0$. The left set of
  runs was performed using the incoming velocity $v_{\rm in}=0$, the
  right set using $v_{\rm in}=2$.  Here $R_{\rm BSS}$, $R_{\rm BB}$,
  and $R_{\rm TS}$ give the value of Rayleigh Chi-squared for the
  binary-single-single, binary-binary, and triple-single results.
  Note that the null hypothesis (consistent with isotropy) is met for
  values $\ll 3$.  For values $\gg 3$ the distribution is evidently
  anisotropic.  Note that in all cases $p<0.05$, indicating that our results are statistically significant.
\label{tab:RayleighIsotropy}
}
\end{center}
\end{table}

To test the anisotropic distribution of the various simulation
results, we checked if the mean vector length is close to zero using the
Rayleigh test for uniformity on the sphere. Under isotropy, the test
statistic $R^2/N$ (where $R$ is the resultant vector length and $N$ is
the number of points) should be $\aplt 3$ (for sufficiently large
$N$). The results of this analysis are presented in
Table \ref{tab:RayleighIsotropy}.

\subsection{The fractality of the initial phase-space}

The distribution of the incoming angles suggest a fractal distribution
in parameter space. For that reason we calculated the fractal (box
counting) dimension ($D$) for each of the distributions (see
Table \ref{tab:RayleighIsotropy}).  To quantify this statement, we present
in Fig. \ref{fig:Mollweide_projection} the fractal dimension by mapping
the azimuth and polar angles on a two-dimensional grid, and
subsequently apply box‑counting to estimate the fractal dimension from
how the number of occupied boxes scales with box size. The results are
presented in Table \ref{tab:FractalDimension}.

\begin{table}
\caption{Fractal dimension for the distributions.}
  \begin{center}
\begin{tabular}{|l|lll|lll|}
\hline
                     & \multicolumn{3}{|c}{$v_{\rm in}=0$} & \multicolumn{3}{|c|}{$v_{\rm in}=2$} \\
Model                & $D_{\rm BSS}$ & $D_{\rm BB}$& $D_{\rm TS}$ & $D_{\rm BSS}$ & $D_{\rm BB}$& $D_{\rm TS}$ \\
\hline
Figure-8             & 1.0 & 0.4 & 0.2  &  1.0 & 0.4 & 0.4 \\
Model A              & 0.9 & 0.2 & 0.2  &  0.8 & 0.6 & 0.1 \\
$I.A.^{i.c.}_4(0.5)$   & 1.2 & 0.1 & 0.3  & 1.1  & 0.5 & 0.4 \\
$I.A.^{i.c.}_{68}(0.5)$ & 0.9 & 0.2 & 0.2  &  0.8 & 0.3 & 0.3 \\
\hline
\end{tabular}

  \tablefoot{Fractal (box counting) dimension for the distributions over
  azimuth and polar angles for the various simulation outcomes for
  braids Figure-8, model A, $I.A.^{i.c.}_4(0.5)$, and
  $I.A.^{i.c.}_{68}(0.5)$ (as in Table \ref{tab:RayleighIsotropy}).  Here
  $D_{\rm BSS}$, $D_{\rm BB}$, and $D_{\rm TS}$ give the box counting
  dimension for the binary-single-single, binary-binary, and
  triple-single results.  Note the value of $D_{\rm ts}$ for model A
  is indicated as an order of magnitude estimate, because the number
  of successful simulations was too small to make a reliable estimate
  of the fractal dimension.
\label{tab:FractalDimension}
}
\end{center}
\end{table}

All binary-single-single encounters correspond to a fractal dimension
$D \sim 1$: the resulting parameter space maps homogeneously on
azimuth and polar angle. We saw this already in relation to
Fig. \ref{fig:Mollweide_projection} (green circles), and the Rayleigh test
results in Table \ref{tab:RayleighIsotropy}.  The binary-binary and
triple-single encounters, however (irrespective of the braid model or the
incoming velocity, $v_{\rm in}$) are clustered or lie along a curve in
azimuth and polar angles, rather than a uniform filling of the
parameter space. Although the amount of data in our experiment is
rather limited, we argue that these clusterings are related to the
chaotic behavior of four-body encounter.  A similar clustering and
patched stability was observed in three-body encounters as islands of
stability in a sea of chaos \citep{2024A&A...689A..24T}.

\subsection{The effect of numerical accuracy}\label{sect:accuracy}

In our experiment we relied on the reversibility of Newton's equations
of motion. However, the numerical method adopted is not precisely
time-reversible.  We did not perform converged solution simulations
because our analysis is strictly of a statistical nature. Small errors
when integrating in one direction may indeed result in the reversible
solution failing to lead to the same initial conditions. In principle,
this would render the numerical result apprehensive
\citep{2018CNSNS..61..160P}.

The interactions, however, are generally short (lasting $<10$ braid
orbits), and the original braids have a long Lyapunov timescale
($\apgt 1000$ orbits).  As a consequence, the chaotic behavior of the
systems hardly has time to cause numerical errors to grow
substantially.  The vast majority of interactions have a relative
energy error $<10^{-5}$, which should suffice for a statistical
scientific interpretation. In some cases, the energy error at the end
of the simulation is larger, but these are not systematically
associated with one of the interactions of interest.  We therefore
consider our results to be robust and not systematically affected by
numerical effects.

\begin{figure}
\center
\includegraphics[width=1\columnwidth]{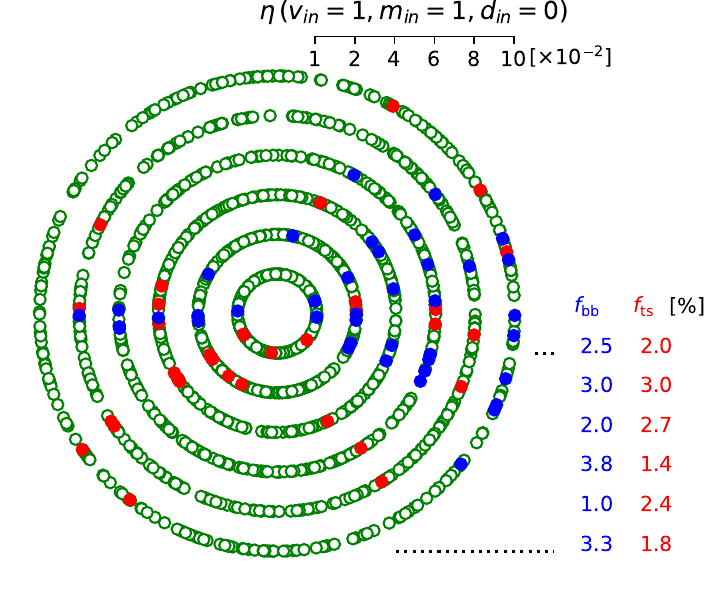}
\caption{ Result of the encounter for the Figure-8 braid for $v_{\rm
    in}=1$, $m_{\rm in}=1$, and $d_{\rm in}=0$ and for various values of
  the numerical time-step parameter ($\eta$).  The $\eta$ tunes the
  time-step of the numerical integration. The default choice in our
  simulations, $\eta=0.01$ (represented by the inner most circle), is
relaxed here for circles farther out, which represent
  less accurate integrations. Each set of runs contains 300 solutions
  for that particular value of $\eta$.
\label{fig:Montgomery_accuracy}
}
\end{figure}

To further support this, we performed an additional series of
simulations based on the Figure-8 braid.  In
Fig. \ref{fig:Montgomery_accuracy} we present the effect of numerical
accuracy on the simulation results. Each circle shows the results (see
the legend and explanation in Fig. \ref{fig:ModelA_results}) for a
series of simulations for a specific value of the time-step parameter
$\eta$, which we varied from 0.01 (inner most circle) to 0.1 (outer
most circle). Not shown is a series of runs for $\eta = 0.001$, but
the results are statistically indistinguishable from $\eta = 0.01$.

The various fractions of simulation outcomes are presented to the
right of the figure (as in Fig. \ref{fig:ModelA_results}).  For this case,
the canonical rate at which a braid forms from two binaries is about
2.6 with a comparable dispersion. Statistically, the probability of a
braid forming through a binary-binary encounter or through a
triple-single encounter are independent of integration accuracy.

\section{Conclusion}

We integrated four different three-body braids, perturbing them with a
fourth particle and evolving until dissolution. Most braids separated
into a binary plus two singles, $\sim 5\%$ separated into two binaries, and
$\sim 3\%$ formed a triple with one escaper.

When the incoming velocity exceeds the braid's binding energy, the
braid may survive temporarily. Symmetric braids dissolve readily under
such strong perturbations, while less symmetric ones may survive if
encountered along the short axis. For instance, braid
$I.A.^{i.c.}_{68}(0.5)$ withstands broadside impacts, whereas model A
breaks into four singles or a binary and two singles.

Reversing ionizing encounters confirms that Figure-8 braids can form from
encounters between two binaries \citep{2000MNRAS.318L..61H}. These
binaries typically have high eccentricities, $\langle e \rangle = 0.80
\pm 0.22$, with a distribution skewness ($S_k \sim -1.26$) to even
higher eccentricities, exceeding those of typical Galactic ($\langle e\ \rangle
\sim 0.6$) and Kuiper belt ($\langle e\ \rangle \sim 0.2$) binaries.\ This
implies that favored initial binary parameters are uncommon.

Braid-formation cross sections for binary--binary ($\sim 5\%$) to
triple--single ($\sim 3\%$) encounters partially offsets the
rareness of suitable binaries. Near-coplanar orbital planes, however,
are required, further restricting formation. Despite this, considering
triples and diverse configurations, braids may be quite common in the
Galaxy \citep{2000MNRAS.318L..61H}.

Although the majority of our calculations were performed in a plane, we
relaxed this assumption by allowing nonplanar interactions. Although
our parameter-space coverage for this generalized approach is limited
to varying a few of the free parameters over a limited range, our
findings show that braids can form a wide range of parameters,
including nonplanar encounters.

We observe that the parameter-space distribution of forming braids is
anisotropic and has a relatively low fractal dimension, forming
distributions in azimuth and polar angles in a two-dimensional space
(much like the earlier work of Jackson Pollock, rather than his late
works). The anisotropy of the initial
parameter space becomes more pronounced for higher incoming
velocities.

Braids are more likely to survive in shallow gravitational potentials,
for example trans-Neptunian objects, the Oort cloud, the Galactic halo, and
intergalactic space. The high Kuiper belt binary fraction is evidence that
transient braids exist in these regions \citep{2008ssbn.book..345N}.

Finite object sizes of astronomical objects suggest braids can form
from collisions during binary--binary or binary--triple
interactions. Stellar braids often lead to collisions unless objects
are compact, suggesting black hole or neutron star braids might form
and be detectable via gravitational waves.  In addition, massive high-velocity runaway stars could form from ionized braids, as one (or
both) of the binaries of the dissolved braid is prone to the two
stars colliding.


\section*{Data availability}

All runscripts and data generated for this paper are available on
Zenodo\url{} under doi 10.5281/zenodo.18403805.  The work was done
with AMUSE, which can be downloaded from {\tt amusecode.org}.

\begin{acknowledgements}
We thank Douglas Heggie, Anna Lisa Varri, and Tjarda Boekholt for
discussing braids. We also thank the referee for the suggestion to
include a discussion on the non-planarity of the problem.  SPZ is
deeply indebted to his friend Joost Visser for joining the expedition
by ferry and train to Edinburgh and attending the Edinburgh Chaotic
Rendez-vous in September 2025.
\end{acknowledgements}

\end{document}